\documentclass{IEEEtran4PSCC}
\ifCLASSINFOpdf
   \usepackage[pdftex]{graphicx}
\else
   \usepackage[dvips]{graphicx}
\fi
\usepackage[cmex10]{amsmath}
\makeatletter
\let\old@ps@headings\ps@headings
\let\old@ps@IEEEtitlepagestyle\ps@IEEEtitlepagestyle
\def\psccfooter#1{%
    \def\ps@headings{%
        \old@ps@headings%
        \def\@oddfoot{\strut\hfill#1\hfill\strut}%
        \def\@evenfoot{\strut\hfill#1\hfill\strut}%
    }%
    \def\ps@IEEEtitlepagestyle{%
        \old@ps@IEEEtitlepagestyle%
        \def\@oddfoot{\strut\hfill#1\hfill\strut}%
        \def\@evenfoot{\strut\hfill#1\hfill\strut}%
    }%
    \ps@headings%
}
\makeatother

\psccfooter{%
        \parbox{\textwidth}{\hrulefill \\ \small{24th Power Systems Computation Conference} \hfill \begin{minipage}{0.2\textwidth}\centering \vspace*{4pt} \includegraphics[scale=0.06]{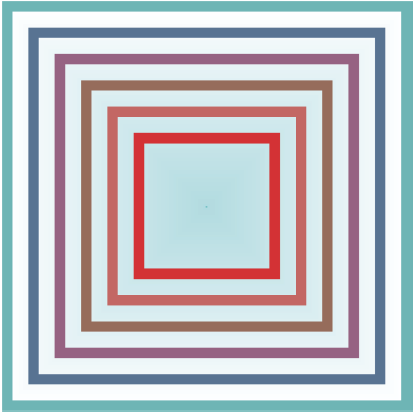}\\\small{PSCC 2026} \end{minipage} \hfill \small{Limassol, Cyprus --- June 8-12, 2026}}%
}

\usepackage{amsmath,amssymb,amsfonts}
\usepackage{algorithmic}
\usepackage{graphicx}
\usepackage{textcomp}
\usepackage[utf8]{inputenc}
\usepackage{xcolor}
\usepackage{orcidlink}
\usepackage{float} 
\usepackage{multirow}
\usepackage{makecell}

\usepackage[linesnumbered,ruled,vlined]{algorithm2e}
\usepackage{booktabs}
\usepackage{cite}

\begin{document}
%
\title{Context-Aware Stochastic Modeling of Consumer Energy Resource Aggregators in Electricity Markets}


\author{
\IEEEauthorblockN{Chatum Sankalpa\textsuperscript{$^{\mathsection,}$\orcidlink{0000-0002-7058-2785}}, Ghulam Mohy-ud-din\textsuperscript{$^{\ddagger,}$\orcidlink{0000-0003-2738-9338}}, Erik Weyer\textsuperscript{$^{\mathsection,}$\orcidlink{0000-0003-4309-4337}} and Maria Vrakopoulou\textsuperscript{$^{\ast,\mathsection,}$\orcidlink{0000-0003-2937-5714}}
}
\IEEEauthorblockA{
$^\mathsection$Electrical and Electronic Engineering, University of Melbourne, Melbourne, Australia \\ 
$^\ddagger$Energy Systems, Energy, Commonwealth Scientific and Industrial Research Organisation, Newcastle, Australia \\
$^\ast$Electrical and Computer Engineering, University of Cyprus, Nicosia, Cyprus \\ 
}
}


\maketitle

\begin{abstract}
Aggregators of consumer energy resources (CERs) like rooftop solar and battery energy storage (BES) face challenges due to their inherent uncertainties. A sensible approach is to use stochastic optimization to handle such uncertainties, which can lead to infeasible problems or loss in revenues if not chosen appropriately. This paper presents three efficient two-stage stochastic optimization methods: risk-neutral, robust, and chance-constrained, to address the impact of CER uncertainties for aggregators who participate in energy and regulation services markets in the Australian National Electricity Market. Furthermore, these methods utilize the flexibility of BES, considering precise state-of-charge dynamics and complementarity constraints, aiming for scalable performance while managing uncertainty. The problems are formed as two-stage stochastic mixed-integer linear programs, with relaxations adopted for large scenario sets. The solution approach employs scenario-based methodologies and affine recourse policies to obtain tractable reformulations. These methods are evaluated across use cases reflecting diverse operational and market settings, uncertainty characteristics, and decision-making preferences, demonstrating their ability to mitigate uncertainty, enhance profitability, and provide context-aware guidance for aggregators in choosing the most appropriate stochastic optimization method.
\end{abstract}

\begin{IEEEkeywords}
aggregator, consumer energy resources, electricity market, stochastic optimization, uncertainty modeling
\end{IEEEkeywords}

\thanksto{\noindent This work is supported partly by the University of Melbourne, the University of Cyprus, and the Department of Climate Change, Energy, the Environment and Water under the International Clean Innovation Researcher Networks (ICIRN) program grant number ICIRN000072, Accelerating Australia’s Power System Transformation -- This project is led by the Commonwealth Scientific and Industrial Research Organisation (CSIRO).}

\vspace{-0.4cm}
\section{Introduction}

\subsection{Motivation}

Australia’s energy transition is accelerating with strong growth in consumer energy resources (CERs) such as rooftop solar photovoltaics (PVs), battery energy storage (BES), loads, and electric vehicles (EVs). Rooftop solar adoption reached a major milestone in December 2024 with the installation of the 4 millionth unit under the ``Small-scale Renewable Energy Scheme", while average system sizes grew from 9.3 kW in Q1 2024 to 9.9 kW in Q1 2025 \cite{DCCEEW2025}. Residential battery uptake is also growing rapidly, with more than 72,000 new installations in 2024 \cite{DCCEEW2025}, representing a 27\% increase from 2023. Looking ahead, major initiatives such as "Integrating Price-Responsive Resources" \cite{AEMC2024} and "Project Jupiter" \cite{ARENA2025} aim to enable large-scale virtual power plant (VPP) participation in Australian electricity markets from 2027 and 2028, respectively, accelerating CER integration and market visibility.

CERs in a VPP platform are integrated into the grid through aggregators, which help manage and coordinate their market participation. 
However, their uncertainties present challenges for maintaining grid stability and reliability. Stochastic optimization provides a structured means to manage these uncertainties \cite{Alonso-Travesset2022}. However, when a specific method is selected without regard to the context of the use case, such as operational and market settings, uncertainty characteristics, and decision-making preferences, it can lead to infeasibilities or revenue losses. In addition, solving stochastic optimization problems with BES poses scalability challenges \cite{Xiao2020} due to time-coupled decisions, complementarity constraints, and the need to co-optimize multiple services, such as energy and reserves, while sharing their flexibility to manage uncertainties. To address these challenges, we propose three stochastic optimization methods for aggregators accounting for CER uncertainties participating in both energy and regulation frequency control ancillary services (FCAS) markets, while offering clear guidance for case-specific method selection.

\subsection{Related Work}

Studies have shown that neglecting CER uncertainty in aggregators’ bidding decisions can cause system balancing issues and lead to revenue losses \cite{Sankalpa2025}. To address this, stochastic optimization methods have been widely applied in aggregator models to capture uncertainties while participating in both energy and reserve markets. Existing research spans CER aggregators across different regions, including Australia \cite{Sankalpa2025, Xiao2020, Ahmad2020, Ghulam2020, Ghulam2022, Najafi-Ghalelou2024}, Europe \cite{Iria2019_B, Iria2019, Iria2019_cluster, Marina2015, Mohammad2023, Wang2023}, and beyond \cite{Xiao2023, Nojavan2024, Ravichandran2018}. In these studies, risk-neutral formulations prioritizing the expected profit remain the most common \cite{Xiao2020, Ghulam2020, Ghulam2022, Najafi-Ghalelou2024, Iria2019_B, Iria2019, Iria2019_cluster, Xiao2023}, although robust \cite{Ahmad2020, Wang2023, Nojavan2024} and chance-constrained \cite{Marina2015,Mohammad2023, Ravichandran2018} methods are increasingly investigated to address operational risk. 

Two-stage risk-neutral stochastic optimization has been used to aggregate residential- and industrial-scale PV, BES, and loads for participation in energy and FCAS markets under PV and load uncertainty \cite{Xiao2020, Ghulam2020}. These studies highlight the role of BES in mitigating uncertainty, but also reveal scalability challenges when managing large-scale aggregations. Similar works replace BES with EVs as flexible resources and use scenario-based methods to solve risk-neutral problems \cite{Iria2019,Iria2019_B,Iria2019_cluster}. These works often permit frequent real-time power imbalances due to uncertainty, whereas this paper, similar to \cite{Sankalpa2025, Najafi-Ghalelou2024, Ahmad2020}, aims to ensure aggregated dispatch conformance (ADC) \cite{AEMO2023}  by keeping bids constant despite uncertainty. Among them, \cite{Ahmad2020} solved a robust optimization problem using affine recourse policies for BES decisions under PV and load uncertainty. In \cite{Nojavan2024}, a robust framework for demand response aggregators in wholesale markets evaluates performance under different uncertainty sets, solving the problem with a column-and-constraint generation method. Chance-constrained scenario-based formulations have been explored in \cite{Marina2015, Mohammad2023, Ravichandran2018} to balance aggregator profitability from CER flexibility in energy markets against the risk of constraint violations under uncertainty. Similar to our work, when solving the chance-constrained problem, \cite{Marina2015} applied the probabilistically robust approach \cite{Roald2014} to a bi-level formulation for a plug-in EV aggregator.

A key gap in the literature is that stochastic methods are often applied without a clear justification for the aggregator’s context, despite variations in operational and market settings, uncertainty characteristics, and decision-making preferences. This motivates our work, which comprehensively compares and evaluates risk-neutral, robust, and chance-constrained approaches for CER aggregators in the NEM using real data, linking method selection to these contexts.

\subsection{Scope and Contributions}

The major contributions of this work are twofold:
\begin{enumerate}
    \item Providing guidance for method selection based on the context of the use case by modeling and comparing two-stage risk-neutral, robust, and chance-constrained stochastic optimization methods that hedge against CER uncertainty, assisting aggregators' efficient participation in energy and regulation reserve markets. As part of modeling, we aim to achieve ADC and problem feasibility in real-time in a probabilistic sense. 
    \item 
    Enhancing the scalability of these stochastic optimization formulations by employing computationally efficient methods that (i) capture BES flexibility through state-of-charge (SoC) dynamics and complementarity constraints, and (ii) reduce the dimensionality of the problem when integrating large numbers of CERs. This contribution complements the first by providing the modeling and computational strategies needed to make method selection practical under uncertainty.
\end{enumerate}

\section{The Proposed Approach}

The NEM is a real-time wholesale electricity market that co-optimizes energy and FCAS requirements for each 5-minute dispatch interval to ensure supply-demand balance. We consider a price-taking aggregator that integrates $b$ BES systems and $p$ solar PV units as flexible resources, along with $d$ inflexible loads, to participate in the NEM’s energy and regulation FCAS (raise and lower) markets. Due to the limited accessibility of network data, we aggregate these resources while neglecting network constraints, assuming that all resources are virtually connected to a single distribution bus. This assumption is reasonable when the resources are geographically located close to each other.

In this paper, the aggregator acts as both a retailer and a supplier by selling electricity to, and purchasing electricity from, CER owners. Moreover, it is assumed that the aggregator signs a contract with CER owners to directly control their flexible assets, such as BES and solar PV units, via a central coordination system to accommodate the dispatch instructions received from the system operator. Since the failure to provide regulation FCAS may lead to contract termination, we assume that the aggregator will force its resources to have adequate capacity to provide contracted FCAS.

In our formulation, the aggregator determines and submits bids for both energy and FCAS for each 5-minute dispatch interval of the next operating day. A full-day horizon is used to decide sensible bids that account for time-coupled BES constraints. During real-time operation, these bids must be met despite uncertainties such as forecast errors; otherwise, the aggregator faces financial penalties for ADC violations and may contribute to system imbalance. To mitigate this, we assume the aggregator can observe uncertainties as they unfold and adjust flexible resources to offset CER forecast deviations while meeting market commitments. The resulting profit-maximization problem under uncertainty is modeled within a two-stage stochastic optimization framework.

The first stage, occurring before the start of the next operating day, is based on PV and load forecasts (i.e., $\{\tilde{P}^{PV,f}_{t}\}_{t \in T} \in \mathbb{R}^{p \times |T|}$ and $\{\tilde{P}^{D,f}_{t}\}_{t \in T} \in \mathbb{R}^{d \times |T|}$) and determines the ``here-and-now" decisions for each 5-minute interval $t \in T$, prior to the realization of uncertainty. These decisions include the energy bid $\{P_{t}^{e}\}_{t \in T} \in \mathbb{R}^{|T|}$, the FCAS raise and lower bids $\{P_{t}^{r}\}_{t \in T}, \{P_{t}^{l}\}_{t \in T} \in \mathbb{R}^{|T|}$, and the corresponding responses of each flexible resource in the aggregator’s portfolio. For BES systems, these decisions comprise the energy market dispatch; charging($c$) $\{P^{B,c}_{t}\}_{t \in T} \in \mathbb{R}^{b \times |T|}$ or discharging($d$) $\{P^{B,d}_{t}\}_{t \in T} \in \mathbb{R}^{b \times |T|}$, together with their FCAS raise($r$)/lower($l$) commitments while charging or discharging $\{P^{B,c,r}_{t}\}_{t \in T}$, $\{P^{B,c,l}_{t}\}_{t \in T}$, $\{P^{B,d,r}_{t}\}_{t \in T}$, $\{P^{B,d,l}_{t}\}_{t \in T}$ $\in \mathbb{R}^{b \times |T|}$, and the state-of-charge (SoC) trajectory $\{E^{B}_{t}\}_{t \in T} \in \mathbb{R}^{b \times |T|}$. Similarly, for flexible PV units, the first-stage decisions include the energy market dispatch $\{P^{PV}_{t}\}_{t \in T} \in \mathbb{R}^{p \times |T|}$ and their FCAS raise/lower commitments $\{P^{PV,r}_{t}\}_{t \in T}$, $\{P^{PV,l}_{t}\}_{t \in T}$ $\in \mathbb{R}^{p \times |T|}$. For compact notation, we collect all these first-stage decision variables in the vector $x \in \mathbb{R}^{(3 + 7b + 3p) \times |T|}$.

The second stage determines the ``wait-and-see” recourse decisions, i.e., the updated operating points of each flexible resource after the realization of uncertainty in every 5-minute interval. These decisions include the real-time BES responses $\{P^{B,c}_{t}(\xi), P^{B,d}_{t}(\xi), P^{B,c,r}_{t}(\xi), \dots , E^B_{t}(\xi)\}_{t \in T} \in \mathbb{R}^{7b \times |T|}$ and PV responses $\{P^{PV}_{t}(\xi), P^{PV,r}_{t}(\xi), P^{PV,l}_{t}(\xi)\}_{t \in T} \in \mathbb{R}^{p \times |T|}$. These updated operating points arise from power adjustments made to compensate for total PV and load forecast errors, $\xi =(\{1_{1\times p} \Delta \tilde{P}^{PV}_{t}\}_{t \in T},  \{1_{1\times d} \Delta \tilde{P}^{D}_{t}\}_{t \in T}) \in \mathbb{R}^{2|T|}$. This recourse mechanism enables the aggregator to follow its bid in real time while mitigating the risk of financial penalties. For compactness, the second-stage decision variables (i.e., the updated BES and PV operating points) are collectively denoted by $y_\xi \in \mathbb{R}^{(7b + 3p) \times |T|}$. 

In case the aggregator fails to meet the energy bid under uncertainty, we introduce slack variables $\{P^{k,-}_{t}{(\xi)}\}_{t \in T}, \{P^{k,+}_{t}{(\xi)}\}_{t \in T} \in \mathbb{R}^{|T|}$ to ensure the power balance. These variables capture the ADC violations arising from negative(-) and positive(+) energy bid imbalances. A negative imbalance $P^{k,-}_{t}{(\xi)}$ is a surplus of demand or a shortage of generation, whereas a positive imbalance $P^{k,+}_{t}{(\xi)}$ is a shortage of demand or a surplus of generation.  For compact notation, we denote these variables by $k^{-}_\xi, k^{+}_\xi \in \mathbb{R}^{|T|}$.

In the subsequent analysis, the two-stage problem is solved under three different formulations: risk-neutral, robust, and chance-constrained.

\subsection{Two-stage Risk-neutral Stochastic Optimization} 

The risk-neutral optimization approach assumes complete knowledge of the probability distributions of uncertain parameters and focuses on maximizing the expected profit. A high-level formulation with compact notations is presented in \eqref{RN1_1}. 
\vspace{-0.5cm}

\begin{subequations}
\label{RN1_1}
\begin{alignat}{1}
\max_{{x}, {y_\xi}, {k^{-}_{\xi}}, {k^{+}_{\xi}}} \ & \Big(f^F(x)+ \mathbb{E}_{\xi}[- \pi^k(k^{-}_{\xi} + k^{+}_{\xi})]\Big), \label{RN1_1a} \\
\textrm{s.t.} \quad 
 & h^F({x}) = 0, \hspace{0.5cm} g^F({x}) \leq 0 \label{RN1_1b}, \\
 & -{k^{-}_{\xi}} \leq h^S(x,{y_\xi},{\xi}) \leq {k^{+}_{\xi}},\hspace{0.66cm} \forall \hspace{0.1cm} {\xi} \label{RN1_1c}, \\
 & {k^{-}_{\xi}},  {k^{+}_{\xi}} \geq 0, \hspace{1cm} \forall \hspace{0.1cm} {\xi} \label{RN1_1d}, \\
 & g^S({y_\xi}, \xi) \leq 0,\hspace{0.54cm} \forall \hspace{0.1cm} {\xi} \label{RN1_1e}. 
\end{alignat}
\end{subequations}

The objective function \eqref{RN1_1a} maximizes both the first-stage and expected second-stage profits. The first term $f^F(x)$ represents revenue from the energy bid and FCAS commitments. The FCAS commitments are paid regardless of whether the regulation is activated. The second term accounts for expected penalty costs, 
$\pi^k(k^{-}_\xi + k^{+}_\xi)$,
incurred for real-time ADC violations due to uncertainty $\xi$.  The first-stage constraints \eqref{RN1_1b} correspond to energy and FCAS power balance equations (equality constraint) and limits of the flexible resources (inequality constraint). The second-stage constraints \eqref{RN1_1c}-\eqref{RN1_1e} should be satisfied for all $\xi$. Constraints \eqref{RN1_1c}-\eqref{RN1_1d} ensure the power balance via non-negative slack variables $k^{-}_\xi$ and $k^{+}_\xi$ which are heavily penalized in the objective function (using a large $\pi^k$) to discourage violations. However, for the FCAS power balance, we set $k^{-}_\xi = k^{+}_\xi = 0$, since any failure to meet FCAS commitments may lead to contract termination. \eqref{RN1_1e} guarantees that the updated operating points of flexible resources used to compensate for uncertainty remain within their limits.

\subsection{Two-stage Robust Optimization} 

The robust optimization approach models uncertainties within a predefined set and determines decisions that remain feasible under worst-case realizations, without requiring probability distributions. A high-level formulation is given in \eqref{RO_1}.
\vspace{-0.5cm}
\begin{subequations}
\label{RO_1}
\begin{alignat}{1}
\max_{{x}, {y_\xi}, {k^{-}_\xi}, {k^{+}_\xi}} \ & \Big(f^F({x})+ \min_{{\xi} \in \Theta} [- \pi^k(k^{-}_\xi + k^{+}_\xi]\Big), \label{RO_1a}\\
\textrm{s.t.} \quad 
 & h^F({x}) = 0, \hspace{0.5cm} g^F({x}) \leq 0, \label{RO_1b} \\ 
 & -k^{-}_\xi \leq h^S(x,{y_\xi}, {\xi}) \leq k^{+}_\xi, \hspace{0.65cm} \forall \hspace{0.1cm} {\xi} \in \Theta, \label{RO_1c}\\
 & k^{-}_\xi, k^{+}_\xi \geq 0, \hspace{1.0cm} \forall \hspace{0.1cm} {\xi}  \in \Theta \label{RO_1d}, \\
 & g^S({y_\xi}, {\xi}) \leq 0,
 \hspace{0.5cm} \forall \hspace{0.1cm} {\xi} \in \Theta \label{RO_1e}.
\end{alignat}
\end{subequations}

The objective function \eqref{RO_1a} maximizes first-stage profit while accounting for the worst-case second-stage outcome over the uncertainty set $\Theta$, which captures all possible realizations of $\xi$. The second-stage constraints \eqref{RO_1c}–\eqref{RO_1e} are required to hold for every $\xi \in \Theta$. As in \eqref{RN1_1}, the second-stage power balance constraint is relaxed.

\subsection{Two-stage Chance-constrained Stochastic Optimization}
\vspace{-0.5cm}
\begin{subequations}
\label{CC_1}
\begin{alignat}{1}
\max_{x, y_\xi} \ \min_{\Theta_{\epsilon}} \ & f^F({x}), \label{CC_1a} \\
\textrm{s.t.} \ 
 & h^F({x}) = 0, \hspace{0.5cm} g^F({x}) \leq 0, \label{CC_1b}\\ 
 & h^S(x, {y_\xi}, {\xi}) = 0, \hspace{0.1cm} 
 g^S({y_\xi},\xi) \leq 0, \hspace{0.1cm} \forall \hspace{0.1cm} {\xi} \in \Theta_{\epsilon}, \label{CC_1c} \\
 & \mathbb{P} \Big({\xi}\in \Theta_{\epsilon}\Big) \geq 1 -\epsilon. \label{CC_1d}
\end{alignat}
\end{subequations}

The chance-constrained formulation in \eqref{CC_1} optimizes the profit over a reduced set $\Theta_{\epsilon} \subset \Theta$, assuming a small probability $\epsilon$ for constraint violation. Due to that relaxation, \eqref{CC_1a} does not contain an ADC violation penalty in this form. The probabilistic constraint \eqref{CC_1d} guarantees that the second-stage constraints \eqref{CC_1c} are satisfied with probability at least 1 - $\epsilon$. 

\subsection{Solution Apporaches}
\label{SC}

Two-stage stochastic optimization problems under continuous uncertainty distributions (e.g., PV and load) are difficult to solve because they are infinite-dimensional with respect to second-stage decisions and constraints. Therefore, to solve the problems efficiently, we employ scenario-based methodologies, representing PV and load forecast errors as scenarios in the second stage. To forecast PV and load, Auto-Regression (AR) models \cite{Sankalpa2022} are estimated from historically observed data. Given point forecasts for target intervals, multiple forecast errors are then generated using a Markov chain Monte Carlo (MCMC) method \cite{mcmc, Vrakopoulou2013}, with the process outlined in Algorithm \ref{alg:mcmc}. It captures temporal correlation by discretizing historical relative forecast errors into states, estimating a transition probability matrix, and sampling scenario trajectories that are finally mapped back to forecast errors and actual values for the target intervals.

Another common approach to simplifying the infinite-dimensional second-stage decisions $y_\xi$ is to approximate them using linear functions of first-stage decisions and uncertain parameters $\xi$, referred to as affine recourse policies \cite{mv_tps13}:
\vspace{-0.1cm}
\begin{equation}
\label{recoure_policies}
y_\xi = y + d^\top \xi.
\end{equation}

where $y$ is the ``here-and-now” part of $y_\xi$ which is made
before the realization of uncertainty while the ``wait-and-see” recourse
term $d^\top \xi$ adjusts itself
to varying data when the uncertain parameters are revealed. This is a good modeling choice when the aggregator has limited ability to re-optimize recourse decisions $y_\xi$ in real time. The coefficient $d$ is also an optimization variable. A suboptimal solution may result from restricting our problems to this affine policy; therefore, we leave it for comparison against the optimal solution.

In problem \eqref{RN1_1}, we use the sample average approximation \cite{Roald2014} to approximate second-stage revenues as 
\begin{equation}
\label{RN1_sol}
\mathbb{E}_{\xi}[- \pi^k({k^-_\xi}+{k^+_\xi})] \approx \sum_{s \in S} p_s \Big(- \pi^k({k^-_{\xi_s}}+{k^+_{\xi_s}})\Big).
\end{equation}
Here $\xi_s$ represents the $s^{th}$ possible outcome of the uncertain parameter $\xi$, and $p_s$ is the associated probability. $S$ is the set of all possible outcomes. We assume a finite scenario set $S$ generated via the MCMC method when solving the problem.

In problem \eqref{RO_1}, we define a box uncertainty set whose bounds are determined by the scenario set $S$. Owing to the linear structure of the robust optimization problem, a tractable reformulation can be obtained by directly identifying the worst-case scenarios located at the boundaries of the uncertainty set. This approach ensures reliability for all realizations within the defined set; however, it may lead to conservative solutions that compromise performance.

To solve the chance-constrained problem in \eqref{CC_1}, we adopt the probabilistically robust approach, a randomized optimization technique employed in \cite{Roald2014}. Unlike the classical scenario approach \cite{Calafiore2006}, this method enables us to consider a substantially reduced number of scenarios within the optimization, while still ensuring that the violation probability does not exceed $\epsilon$. The procedure consists of two main steps. In the first step, with confidence level at least $1-\beta$, we identify the minimum-volume uncertainty set $\Theta^{'}_{\epsilon}$ that contains at least $1-\epsilon$ of the probability mass of the uncertain parameter $\xi$. The number of sampled scenarios required for this step depends on the dimension of uncertainty $N_\xi$ and is given by \eqref{scenarios}:
\vspace{-0.1cm}
\begin{equation}
\label{scenarios}
 N_s \geq \frac{1}{\epsilon} \frac{e}{e-1}\left(\ln \frac{1}{\beta} + N_\xi -1 \right).
\end{equation}

where $e$ denotes Euler's number. In the second step, the probabilistically constructed set $\Theta^{'}_{\epsilon}$  is employed to reformulate the chance constraints \eqref{CC_1c}-\eqref{CC_1d} as robust constraints, ensuring the feasibility for all realizations within $\Theta^{'}_{\epsilon}$:
\vspace{-0.1cm}
\begin{equation}
\label{CC_RO} 
 h^S(x, {y_\xi}, {\xi}) = 0, \hspace{0.1cm} 
 g^S({y_\xi},\xi) \leq 0, \hspace{0.4cm} \forall \hspace{0.1cm} {\xi} \in \Theta^{'}_{\epsilon}.
\end{equation}

This approach guarantees that any feasible solution to robust reformulation \eqref{CC_RO} is feasible for the chance constraint with confidence at least $1-\beta$. However, to improve feasibility, the power balance constraint in \eqref{CC_RO} can be relaxed, similar to \eqref{RO_1}. 

\begin{algorithm}[htbp]
\caption{Scenario Generation Using MCMC} 
\label{alg:mcmc}
\KwIn{
    Historical forecast errors $\{\epsilon_{t_h}\}_{t_h=1}^{T_h}$ \\
    Point forecasts for historical intervals $\{\hat{P}_{t_h}\}_{t_h=1}^{T_h}$ \\
    Point forecasts for target intervals $\{\hat{P}_t\}_{t=1}^{T}$ \\
    $S$ discrete states, $N_{\text{scen}}$ scenarios, 
    $T$ target intervals 
}
\KwOut{
    Scenario trajectories $\{P_t^{(n)}\}_{t=1}^{T}, \; n = 1, \dots, N_{\text{scen}}$
}

\textbf{Step 1: Data Preprocessing} \\
Normalize historical forecast errors to relative errors: \\
\For{$t_h = 1$ \KwTo $T_h$}{
    $r_{t_h} \leftarrow \epsilon_{t_h} / \hat{P}_{t_h}$
}

\textbf{Step 2: Discretization and State Assignment} \\
Partition the range of $r_{t_h}$ into $S$ non-overlapping bins: $B_s = [b_{s-1}, b_s)$ for $s = 1, 2, \dots, S{-}1$ and $B_S = [b_{S-1}, b_S]$\\
Assign each $r_{t_h}$ to a discrete state $s_{t_h} \in \{1, 2, \dots, S\}$ such that $r_{t_h} \in B_{s}$\\
Record bin centers $c_s = \frac{b_{s-1} + b_s}{2}$ for each state $s = 1, \dots, S$ \\

\textbf{Step 3: Transition Probability Matrix Estimation} \\
Estimate first-order Markov transition matrix $\Pi \in \mathbb{R}^{S \times S}$: \\
$\Pi_{ij} = \mathbb{P}(s_{t_h+1} = j \mid s_{t_h} = i)$

\textbf{Step 4: Scenario Sampling via MCMC} \\
\For{$n = 1$ \KwTo $N_{\text{scen}}$}{
    Initialize $s^{(n)}_1$ randomly based on empirical state distribution. \\
    \For{$t = 2$ \KwTo $T$}{
        Sample $s^{(n)}_t \sim \Pi_{s^{(n)}_{t-1}, :}$ \\
    }
    Obtain relative error sequence: $r^{(n)}_t \leftarrow c_{s^{(n)}_t}$
}

\textbf{Step 5: Mapping to Forecast Errors and Actual Values} \\
\For{$n = 1$ \KwTo $N_{\text{scen}}$}{
    \For{$t = 1$ \KwTo $T$}{
        Forecast error: $\epsilon^{(n)}_t = r^{(n)}_t \cdot \hat{P}_t$ \\
        Actual value: $P^{(n)}_t = \hat{P}_t + \epsilon^{(n)}_t$
    }
}
\end{algorithm}

\subsection{Detailed Risk-neutral Scenario-based Reformulation}

We now detail the scenario-based reformulation \eqref{RN_1}, which is used to obtain a tractable solution to the high-level risk-neutral formulation in \eqref{RN1_1}. Here, the dependence of the second-stage variables on the uncertain parameter $\xi$ is replaced by an explicit scenario index $s \in S$.
In the objective function \eqref{RN_1a}, the first term captures the net revenue from energy bids and FCAS commitments, where $\pi^e_t, \pi^r_t, \pi^l_t \in \mathbb{R}$ represent the market clearing prices (MCPs) for energy, regulation FCAS raise and lower services, respectively at time $t$. The second term reflects the expected penalty costs due to ADC violations leading from real-time energy bid imbalances. 
The parameter $\Delta \tau$ denotes the duration of the optimization interval. We collect the decision variables in the vector $v = [x, y_{\xi}, k^-_\xi, k^+_\xi]^\top \in \mathbb{R}^{(5 + 14b + 6p) \times |T|}$ for compact notation.
\vspace{-0.1cm}
\begin{subequations}
\label{RN_1}
\allowdisplaybreaks
\begin{alignat}{1}
\max_{v} \ & \sum_{t \in T} \Big((\pi^{e}_t P_{t}^{e} + \pi^r_t P_{t}^{r} + \pi^l_t P_{t}^{l}) \Delta \tau \nonumber \\
& \hspace{2.0cm} - \pi^k \sum_{s\in S} p_s (P^{k,-}_{t,s} + P^{k,+}_{t,s})\Big) ,\label{RN_1a}\\
\textrm{s.t.} \quad & \text{First-stage power balance ($\forall \hspace{0.1cm} t \in T$)}, \nonumber \\ 
& 1_{1\times b}(P^{B,d}_{t}-P^{B,c}_{t}) + 1_{1\times p}P^{PV}_{t}  \nonumber \\
& \hspace{3.0cm} - 1_{1\times d}\tilde{P}^{D,f}_{t} = P_{t}^{e},\label{RN_1b}\\
& 1_{1\times b}(P^{B,d,r}_{t} +  P^{B,c,r}_{t}) + 1_{1\times p} P^{PV,r}_{t}= P_{t}^{r} ,\label{RN_1c}\\
& 1_{1\times b}(P^{B,c,l}_{t} + P^{B,d,l}_{t}) + 1_{1\times p} P^{PV,l}_{t} = P_{t}^{l},\label{RN_1d}\\
& \text{First-stage PV and BES constraints ($\forall \hspace{0.1cm} t \in T$)}, \nonumber \\ 
& \text{\eqref{RN_1j}-\eqref{RN_1w} by dropping subscript $s$ ($\xi$ = 0)} ,\label{RN_1e} \\
& \text{Second-stage power balance ($\forall \hspace{0.1cm} s \in S, \hspace{0.1cm} t \in T$)}, \nonumber \\ 
& -P^{k,-}_{t,s} \leq 1_{1\times b}(P^{B,d}_{t,s}-P^{B,c}_{t,s}) + 1_{1\times p}P^{PV}_{t,s} \nonumber \\
& \quad \quad - 1_{1\times d}(\tilde{P}^{D,f}_{t} + \Delta \tilde{P}^{D}_{t,s}) - P_{t}^{e} \leq P^{k,+}_{t,s} ,\label{RN_1f}\\
& P^{k,+}_{t,s}, P^{k,-}_{t,s} \geq 0 ,\label{RN_1g}\\
& 1_{1\times b}(P^{B,d,r}_{t,s} +  P^{B,c,r}_{t,s}) + 1_{1\times p} P^{PV,r}_{t,s}= P_{t}^{r} ,\label{RN_1h}\\
& 1_{1\times b}(P^{B,c,l}_{t,s} + P^{B,d,l}_{t,s}) + 1_{1\times p} P^{PV,l}_{t,s} = P_{t}^{l},\label{RN_1i}\\
& \text{Second-stage PV constraints ($\forall \hspace{0.1cm} s \in S, \hspace{0.1cm} t \in T$)}, \nonumber \\ 
& 0 \leq P^{PV,r}_{t,s} \leq  (\tilde{P}^{PV,f}_{t} + \Delta \tilde{P}^{PV}_{t,s}) - P^{PV}_{t,s} ,\label{RN_1j}\\
& 0 \leq P^{PV,l}_{t,s} \leq P^{PV}_{t,s} ,\label{RN_1k}\\ 
& \text{Second-stage BES constraints ($\forall \hspace{0.1cm} s \in S, \hspace{0.1cm} t \in T$)}, \nonumber \\ 
& \underline E^{B} \leq E^{B}_{t,s} \leq \overline E^{B} ,\label{RN_1l}\\
& 0 \leq  P^{B,d}_{t,s} \leq \alpha^B_{t,s} \overline P^{B,d} ,\label{RN_1m}\\
& 0 \leq  P^{B,c}_{t,s} \leq (1-\alpha^B_{t,s}) \overline P^{B,c} ,\label{RN_1n}\\
& E^B_{t,s} = E^B_{t-1,s} + (\eta^c P^{B,c}_{t,s} - \frac{1}{\eta^d} P^{B,d}_{t,s}) \Delta \tau  ,\label{RN_1o}\\
& E^B_{0,s} = E^B_{in} ,\label{RN_1p}\\
& E^B_{|T|,s} = E^B_{in} ,\label{RN_1q}\\
& 0 \leq  P^{B,d,r}_{t,s} \leq \overline P^{B,d} -  P^{B,d}_{t,s} ,\label{RN_1r}\\
& 0 \leq  P^{B,c,r}_{t,s} \leq P^{B,c}_{t,s}  ,\label{RN_1s}\\
& 0 \leq  P^{B,c,l}_{t,s} \leq \overline P^{B,c} - P^{B,c}_{t,s},\label{RN_1t}\\
& 0 \leq  P^{B,d,l}_{t,s} \leq P^{B,d}_{t,s} ,\label{RN_1u}\\
& \underline E^{B} \leq E^B_{t,s} - \Big( \frac{P^{B,d,r}_{t,s}}{\eta^d} + P^{B,c,r}_{t,s}\eta^c \Big) \Delta \tau   ,\label{RN_1v}\\
& E^B_{t,s} + \Big(P^{B,c,l}_{t,s}\eta^c + \frac{P^{B,d,l}_{t,s}}{\eta^d}\Big) \Delta \tau \leq \overline E^{B}. \label{RN_1w}
\end{alignat}
\end{subequations}
\vspace{-0.1cm}
Constraints \eqref{RN_1b}–\eqref{RN_1d} define the first-stage active power balance for energy, FCAS raise, and FCAS lower, based on PV and load forecasts. Constraint \eqref{RN_1e} encompasses all the first-stage PV- and BES-related constraints, which are detailed later when presenting the second-stage formulation.

Constraints  \eqref{RN_1f}-\eqref{RN_1i} define the second-stage active power balance for delivering energy, FCAS raise, and lower in real time under PV and load uncertainty.
Among them, \eqref{RN_1f}-\eqref{RN_1g} ensures the real-time balance of the energy market bid under uncertainty via slack variables, $P^{k,+}_{t,s}$ and $P^{k,-}_{t,s}$. Since failure to provide FCAS may result in contract termination, the aggregator must control its resources to provide the contracted regulation FCAS raise and lower services in real time, regardless of uncertainty, as given by constraints \eqref{RN_1h}-\eqref{RN_1i}. 

Constraints \eqref{RN_1j}-\eqref{RN_1k} define the second-stage PV limits. \eqref{RN_1j} ensures that FCAS raise from PV is non-negative and bounded by the remaining PV capacity after part of the PV output is allocated to the energy market, while \eqref{RN_1k} ensures that FCAS lower from PV is non-negative and bounded by the PV generation committed to the energy market.

Constraints \eqref{RN_1l}-\eqref{RN_1w} define the second-stage BES operating limits. \eqref{RN_1l} restricts the SoC within its minimum and maximum capacity bounds, $\underline E^{B}, \overline E^{B} \in \mathbb{R}^b$. To reduce computational complexity and enforce mutual exclusivity between charging and discharging, the typical nonlinear and nonconvex complementarity condition of the BES,   $P^{B,c}_{t,s}.P^{B,d}_{t,s} = 0$, is reformulated using the Fortuny-Amat transformation \cite{Fortuny-Amat1981}. This introduces binary variables $\alpha^B_{t,s} \in \{0,1\}$ in a mixed-integer linear programming (MILP) framework, yielding constraints \eqref{RN_1m}-\eqref{RN_1n}. The parameters $\overline P^{B,c}, \overline P^{B,d} \in \mathbb{R}^b$ denote maximum charging and discharging capacities and serve as tight big-$M$ constants. This exact MILP formulation ensures that only one action; charging or discharging, can occur at a time.

The SoC evolution of BES after energy market deployment is captured by \eqref{RN_1o}. Initial and terminal SoC conditions are enforced by \eqref{RN_1p}-\eqref{RN_1q}, requiring the BES to return to its initial value $E^{B}_{in} \in \mathbb{R}$ at the end of the horizon.

The FCAS provision of BES is modeled as follows: raise services can be supplied by either increasing discharging or reducing charging, with corresponding limits defined in \eqref{RN_1r}-\eqref{RN_1s}; lower services can be supplied by either increasing charging or reducing discharging, with limits given in \eqref{RN_1t}-\eqref{RN_1u}. Because raise actions reduce stored energy, constraint \eqref{RN_1v} prevents the SoC from falling below $\underline E^{B}$. Conversely, lower actions increase stored energy, and constraint \eqref{RN_1w} ensures that SoC does not exceed $\overline E^{B}$.

The problem is first addressed using the scenario-based method discussed in Section \ref{SC}. For comparison, we also re-solve it employing affine recourse policies. For example, the second-stage BES charging decision based on the affine recourse policy \eqref{recoure_policies} is given by,
\vspace{-0.1cm}
\begin{equation}
P^{B,c}_{t,s} = P^{B,c}_{t} + d^{B,c}_{PV,t} (1_{1\times p} \Delta \tilde{P}^{PV}_{t}) + d^{B,c}_{D,t} (1_{1\times d} \Delta \tilde{P}^{D}_{t}). \label{Affine_recoure_policies}
\end{equation}

where $P^{B,c}_{t} \in \mathbb{R}^b$ is the first-stage counterpart of $P^{B,c}_{t,s} \in \mathbb{R}^b$, and $d^{B,c}_{PV,t}$, $d^{B,c}_{PV,t} \in \mathbb{R}^b$ are variables that adjust power in real time, compensating for PV and load forecast errors. 

\subsection{Improving Scalability under Large Scenario Sets}
\label{scalable}

The Fortuny-Amat transformation employed to reformulate the nonlinear complementarity condition of the BES, $P^{B,c}_{t,s}.P^{B,d}_{t,s} = 0$, is not scalable in the risk-neutral stochastic optimization setting with a large scenario set $S$, as the inclusion of additional binary variables substantially enlarges the search space and slows down the solution process. To reduce computational complexity and improve scalability, we relax the MILP to a linear program (LP) by removing the binary variables $\alpha^B_{t,s}$ from \eqref{RN_1m}-\eqref{RN_1n}. The relaxed formulation retains \eqref{RN_relax1}-\eqref{RN_relax2}, while introducing a penalty term, $-\lambda \sum_{t\in T} \sum_{s\in S} (P^{B,c}_{t,s} + P^{B,d}_{t,s})$, in the objective function to discourage simultaneous charging and discharging.
\vspace{0.1cm}
\begin{subequations}
\label{RN_relax}
\allowdisplaybreaks
\begin{alignat}{1}
    & 0 \leq  P^{B,d}_{t,s} \leq \overline P^{B,d} , \hspace{0.4cm}
    0 \leq  P^{B,c}_{t,s} \leq  \overline P^{B,c} ,\label{RN_relax1} \\
    & P^{B,c}_{t,s} + P^{B,d}_{t,s} \leq  \text{max}(\overline P^{B,d}, \overline P^{B,c}) \label{RN_relax2}
\end{alignat}
\end{subequations}  

Note that a large $\lambda$ drives both charge and discharge decisions to zero, whereas a small $\lambda$ leads to frequent simultaneous charge/discharge decisions. Therefore, selecting a suitable value for $\lambda$ is crucial for obtaining a reliable solution that is close to the MILP.

In addition to the LP-relaxed formulation described earlier, for comparison, we also consider a more generic convex relaxation approach using McCormick envelopes \cite{McCormick1976}. Here, the bilinear complementarity condition $P^{B,c}_{t,s}.P^{B,d}_{t,s} = 0$ is reformulated by introducing an auxiliary variable $w = P^{B,c}_{t,s}.P^{B,d}_{t,s}$ and replacing the bilinear term with its convex envelope \eqref{ME1}-\eqref{ME2}, over known bounds. Note that this convex relaxation also does not strictly enforce mutual exclusivity.
\begin{subequations}
\label{McCorcmick Envelopes}
\allowdisplaybreaks
\begin{alignat}{1}
    & w \geq  0, \hspace{0.3cm} w \geq  P^{B,c}_{t,s}. \overline P^{B,d} +  P^{B,d}_{t,s}. \overline P^{B,c} - \overline P^{B,c}. \overline P^{B,d}, \label{ME1} \\
    & w \leq  P^{B,c}_{t,s}. \overline P^{B,d}, \hspace{0.3cm} w \leq  P^{B,d}_{t,s}. \overline P^{B,c}. \label{ME2}
\end{alignat}
\end{subequations}

The performance of each reformulation under the risk-neutral setting is compared in Section \ref{results}.

\section{Numerical Experiments and Validation}

We test the proposed stochastic optimization methods on aggregators comprising rooftop PVs, loads, and BES. Experiments are organized along three contexts: (i) operational and market settings (CER capacities, MCPs, violation penalties, and market preferences); (ii) uncertainty characteristics (real-time forecast error levels and uncertainty-set bounds); and (iii) decision-making preferences (trade-off between accurate versus faster suboptimal decisions). In addition to the three stochastic approaches, we include a deterministic benchmark based on point forecasts to quantify the benefits of stochastic modeling. 
A baseline case is defined with all parameters at their original values. Multiple use cases are then generated by varying one or more contextual factors. Performance is compared in terms of profit, computational time, and ADC violations to provide systematic guidance for method selection.

\subsection{Baseline Case}

\begin{figure}[H]
\centerline{\includegraphics[width=0.5\textwidth, trim=0.1cm 0.1cm 0.1cm 0.1cm, clip]{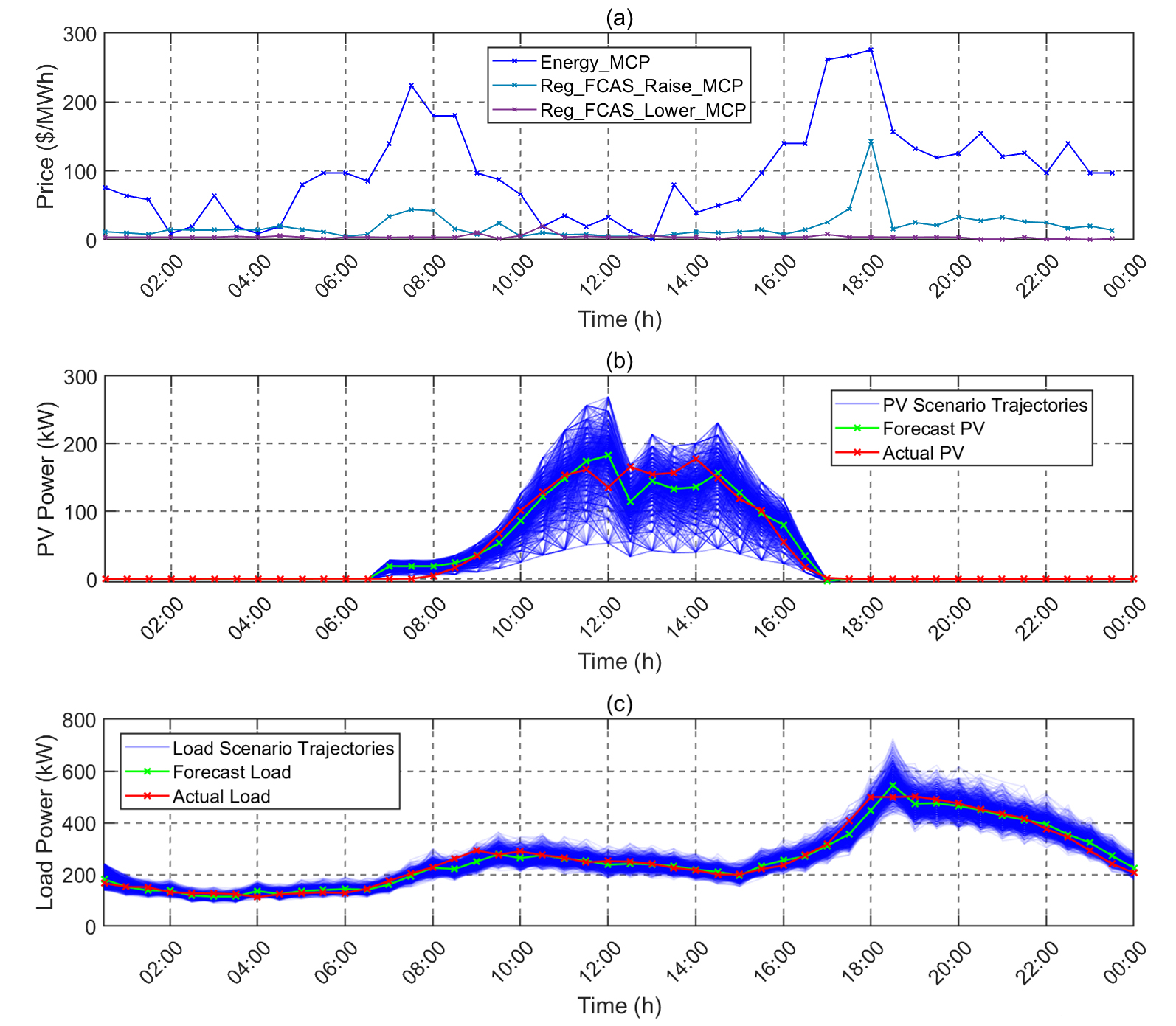}}
\caption{(a) Market clearing prices (MCPs) for energy, regulation FCAS raise, and regulation FCAS lower, (b) Actual and forecast PV values, and 1000 PV scenario trajectories stemming from MCMC-based forecast errors (c) Actual and forecast load values, and 1000 load scenario trajectories stemming from MCMC-based forecast errors, for day 01/07/2012.}
\label{Data_Plots}
\end{figure}

The baseline case considers an aggregator participating in the energy and regulation FCAS markets of the NEM on day 01/07/2012. The aggregator’s portfolio comprises 300 geographically proximate houses from the Ausgrid network, each equipped with a PV unit, a load, and a BES. Three years of 30-minute resolution data (01/07/2010-30/06/2013) are sourced from \cite{Data.NSW}, while market data (energy and regulation FCAS MCPs) are obtained from \cite{AEMO}. For this study, the optimization framework is implemented in a 30-minute interval instead of the standard 5-minute dispatch interval of NEM, although the methods are applicable to any time step. Fig.~\ref{Data_Plots} illustrates the MCPs and uncertainty-related data for day 01/07/2012. The aggregated system parameters are listed in Table
\ref{tab:group_values}.

\begin{table}[H]
\caption{Parameters of the Test System}
\centering
\begin{tabular}{c c c c}
\hline
\textbf{Parameter} & \textbf{Value} & \textbf{Parameter} & \textbf{Value} \\
\hline 
$\overline P^{B,c}, \overline P^{B,d}$ & 5 kW, 5 kW & $\overline E^{B}, E^B_{in}$ & 13.5 kWh, 6.75 kWh \\
$\eta^c, \eta^d$ & 0.95, 0.95 & $\pi^k, \lambda $ & 1 \$/kW , $10^{-7}$ \$/kW \\
\hline
\end{tabular}
\label{tab:group_values}
\end{table}
\vspace{-0.5cm}
\vspace{-0.5cm}
\begin{figure}[H]
\centerline{\includegraphics[width=0.4\textwidth, trim=0.1cm 0.1cm 0.1cm 0.1cm, clip]{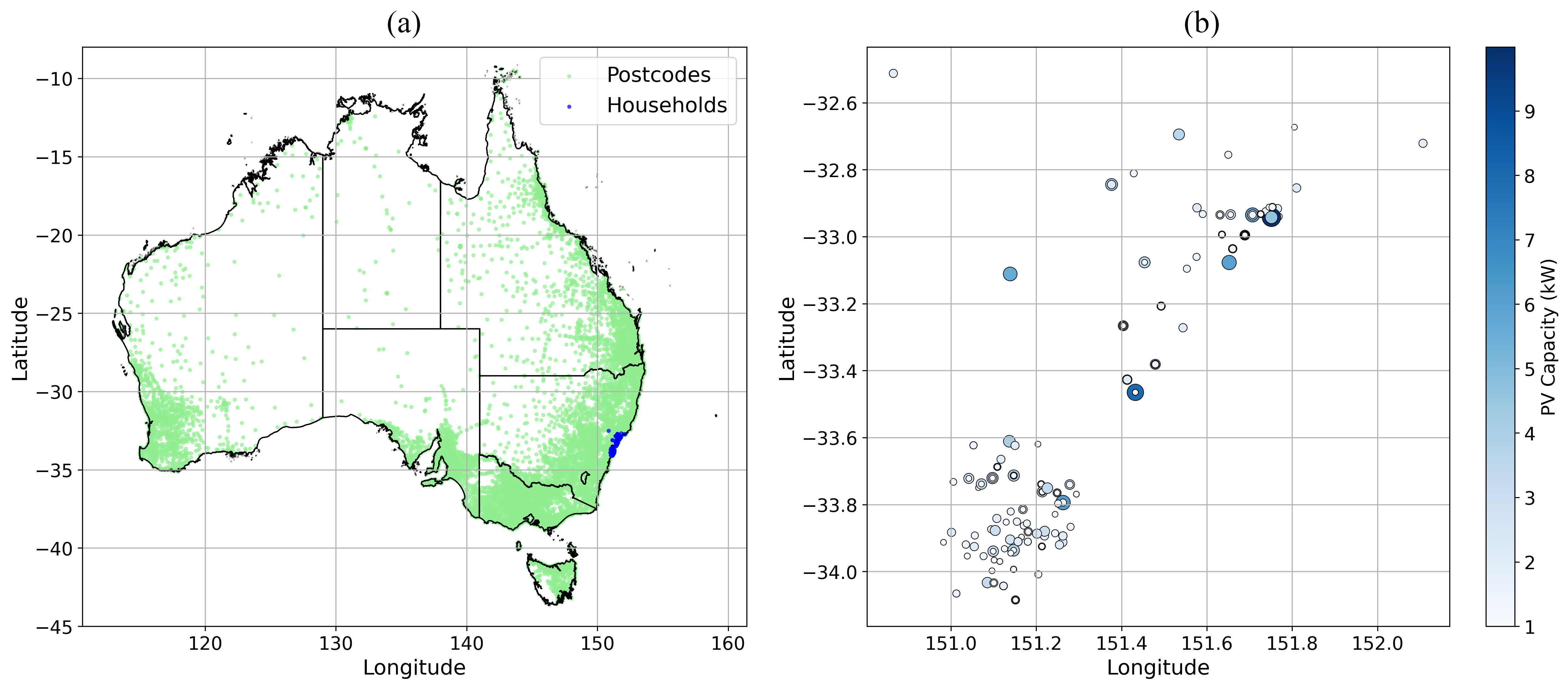}}
\caption{Locations of the (a) houses (in blue) and (b) PV units colored by the installed capacity.}
\label{Locations}
\end{figure}
\vspace{-0.5cm}
\begin{figure}[htbp]
\centerline{\includegraphics[width=0.4\textwidth, trim=0.1cm 0.1cm 0.1cm 0.1cm, clip]{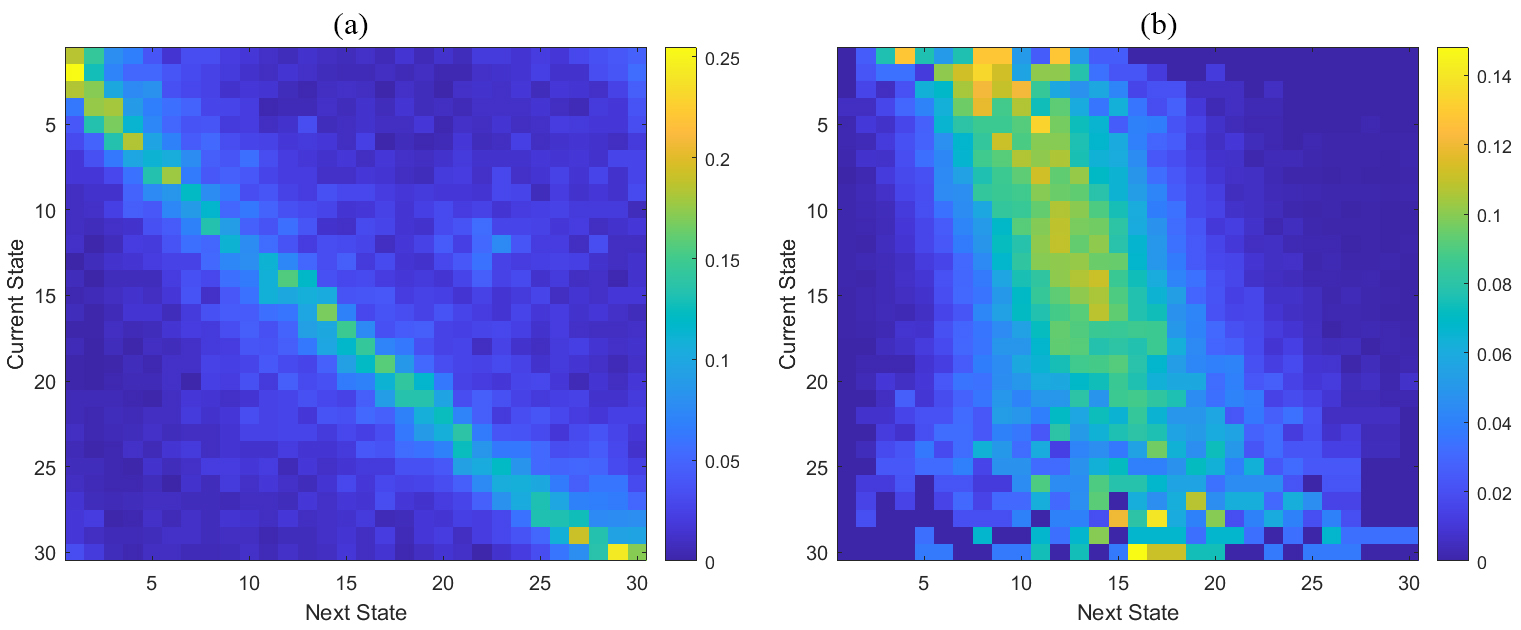}}
\caption{State transition probability matrices for (a) PV forecast errors and (b) load forecast errors. ``Yellow" indicates high probabilities, while ``blue" indicates low probabilities.)}
\label{TPMs}
\end{figure}

The geographical locations of the houses where the CERs are connected, along with the installed capacities of their PV units, are shown in Fig.~\ref{Locations}. As the houses are located in close proximity, network constraints can be reasonably neglected. Hence, PV units and loads are modeled in aggregate rather than individually \cite{Xiao2020}, reflecting their spatial proximity and correlated behavior. Furthermore, all BES systems are assumed to be operationally identical in all parameters, allowing aggregation into a single equivalent BES without feasibility or optimality issues \cite{Xiao2020}. This aggregation preserves the essential flexibility characteristics of the system while reducing problem dimensionality, thereby enhancing both computational tractability and scalability for stochastic optimization. It is noted, however, that if BES systems differ in their operating characteristics, such aggregation may lead to suboptimal or infeasible solutions upon disaggregation, which in this study is performed proportionally to unit capacities.

When modeling uncertainty, we apply the MCMC method described in Algorithm~\ref{alg:mcmc}. First, transition probability matrices are trained using historical relative forecast errors of PV and load for each 30-minute interval over the two years (01/07/2010-30/06/2012). The resulting 30-state transition probability matrices in Fig.~\ref{TPMs} exhibit higher probabilities concentrated around the diagonals, indicating temporal correlation in forecast errors. Using these matrices, we generate multiple PV and load scenario trajectories (1,000 in our case) for day 01/07/2012, as shown in Fig.~\ref{Data_Plots}(b) and Fig.~\ref{Data_Plots}(c).

These second-stage scenarios, together with the corresponding day-ahead point forecasts for the first stage, are used to solve the two-stage stochastic optimization problems for 01/07/2012, which we refer to as the ``training" phase. The performance of the proposed methods is then evaluated against the actual realizations for 01/07/2012, in the ``testing" phase.

We first present numerical results for the baseline case, followed by results from experiments conducted for various use cases within the discussed contexts. Simulations were performed in Julia using Gurobi solver on an Intel\textsuperscript{\textregistered}
Core\textsuperscript{\texttrademark} i7-1265U 1.8 GHz processor with 16 GB RAM. 

\begin{table*}[htbp]
\caption{Use Cases of Different Contexts}
\centering
\begin{tabular}{c c c c c c c c c c c c}
\toprule
\multirow{2}{*}{\textbf{Case}} & \multicolumn{2}{c}{Forecast} & \multicolumn{2}{c}{Uncertainty} & \multirow{2}{*}{\makecell{BES\\Capacity}} & \multirow{2}{*}{\makecell{Energy\\Price}} & \multirow{2}{*}{\makecell{FCAS\\Prices}} & \multirow{2}{*}{\makecell{Penalty\\Cost}} & \multicolumn{2}{c}{Actual} & \multirow{2}{*}{\makecell{Market\\Preference}} \\ 
\cmidrule(lr){2-3} \cmidrule(lr){4-5} \cmidrule(lr){10-11}
 & PV & Load & PV & Load &  &  &  &  & PV & Load &  \\ 
\midrule
\textbf{1}  & Mod & Mod & Mod & Mod & Mod & Mod & Mod & Mod & Mod & Mod & Energy + FCAS \\
\textbf{2}  & Mod & Mod & Mod & Mod & Mod & Mod & Mod & Mod & Mod & Mod & Energy \\
\textbf{3}  & Mod & Mod & Mod & Mod & Mod & High & Low & Mod & Mod & Mod & Energy + FCAS \\
\textbf{4}  & Mod & Mod & Mod & Mod & Mod & Low & High & Mod & Mod & Mod & Energy + FCAS \\
\textbf{5}  & High & Low & High & Low & Mod & Mod & Mod & Mod & High & Low & Energy + FCAS \\
\textbf{6}  & Low & High & Low & High & High & Mod & Mod & Mod & Low & High & Energy + FCAS \\
\textbf{7}  & Mod & Mod & Mod & Mod & High & Mod & Mod & Mod & Mod & Mod & Energy + FCAS \\
\textbf{8}  & High & High & High & High & Mod & Mod & Mod & Mod & High & High & Energy + FCAS \\
\textbf{9}  & High & High & High & High & Low & Mod & Mod & Mod & High & High & Energy + FCAS \\
\textbf{10} & Low & High & Low & High & Low & Mod & Mod & Mod & Low & High & Energy + FCAS \\
\textbf{11} & Mod & Mod & Mod & Mod & Low & Mod & Mod & Mod & Mod & Mod & Energy + FCAS \\
\textbf{12} & Mod & Mod & Mod & Mod & Mod & Mod & Mod & Mod & Mod & 1.25$\times$Mod & Energy + FCAS \\
\textbf{13} & Mod & Mod & Mod & Mod & Mod & Mod & Mod & Low & Mod & 1.25$\times$Mod & Energy + FCAS \\
\textbf{14} & Mod & Mod & Mod & Mod & Mod & Mod & Mod & Mod & 0.5$\times$Low & Mod & Energy + FCAS \\
\textbf{15} & Mod & Mod & 0.5$\times$Low & 0.5$\times$Low & Mod & Mod & Mod & Mod & Mod & Mod & Energy + FCAS \\
\bottomrule
\end{tabular}
\label{context1}
\end{table*}

\subsection{Results and Discussion}
\label{results}

\begin{figure}[htbp]
\centerline{\includegraphics[width=0.51\textwidth, trim=0.8cm 0.1cm 0.9cm 0.1cm, clip]{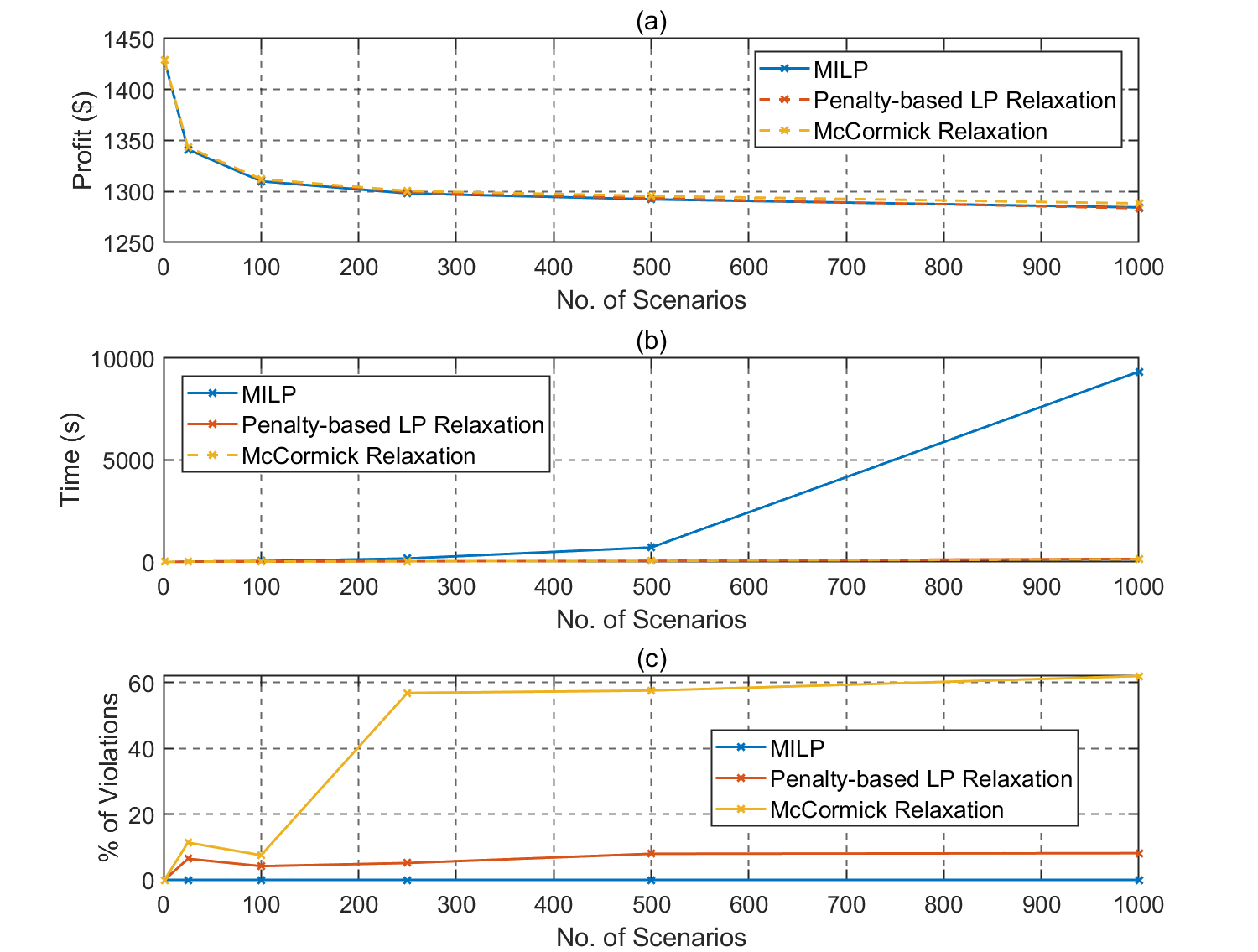}}
\caption{Comparison of (a) profits, (b) computational times, and (c) percentage of mutual exclusivity violations of BES decisions, for different reformulations under risk-neutral setting with increasing number of scenarios.}
\label{profits_time_violations}
\end{figure}

To solve the risk-neutral problem in \eqref{RN1_1}, we used the 1,000 scenario trajectories generated via the MCMC method as the scenario set $S$. However, solving the problem with such a large number of scenarios is computationally demanding when using the MILP reformulation of the BES complementarity condition. We compare the MILP results with the proposed scalable approaches, namely the penalty-based LP relaxation and the McCormick relaxation, under the risk-neutral setting, as discussed in Section~\ref{scalable}. The comparison is made in terms of profit, computational time, and violations of the BES complementarity condition, as illustrated in Fig.~\ref{profits_time_violations}.

\begin{figure}[htbp]
\centerline{\includegraphics[width=0.5\textwidth, trim=1.5cm 0.01cm 2.5cm 0.2cm, clip]{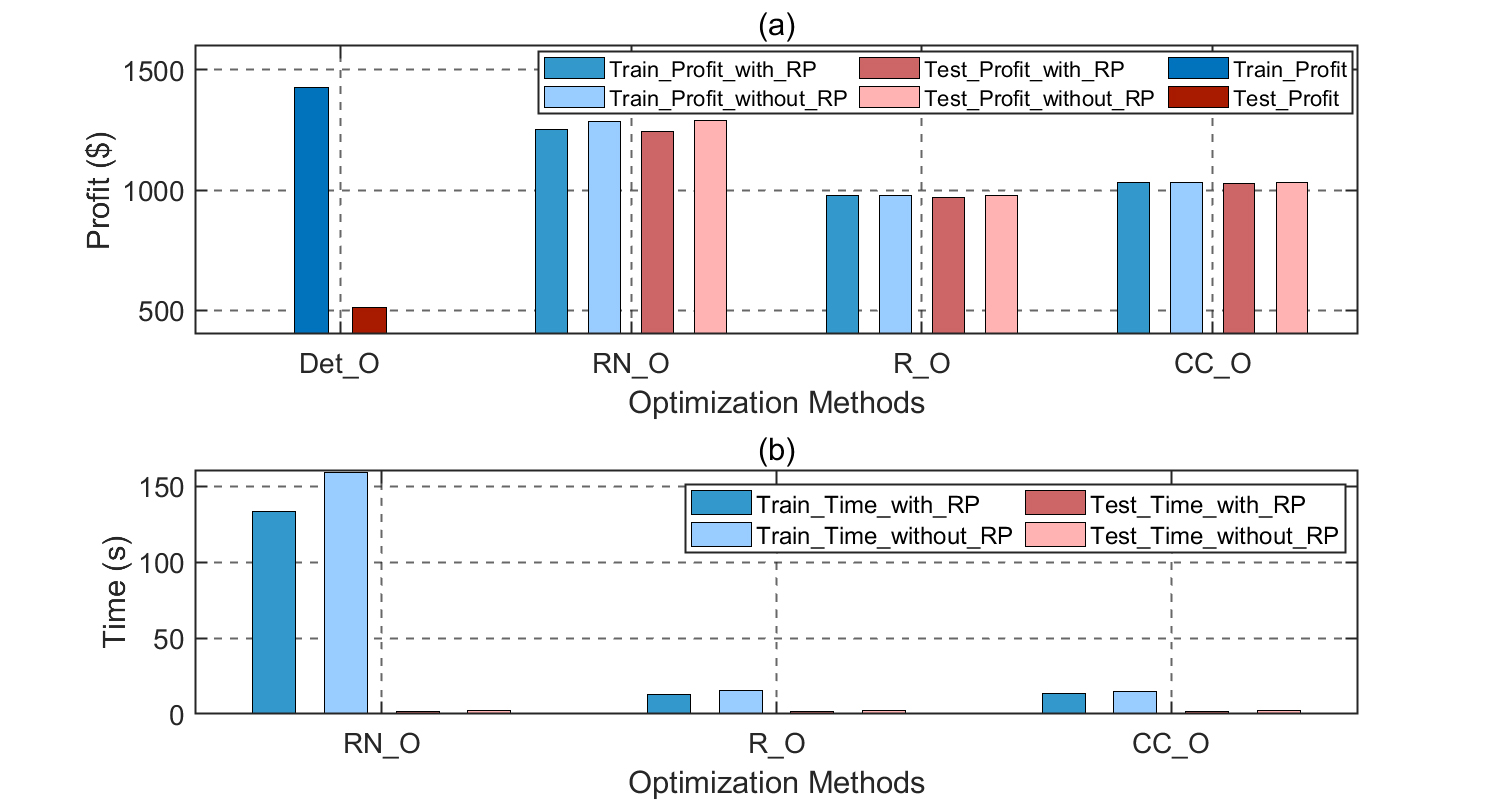}}
\caption{Training and test (a) profits and (b) computational times with and without recourse policies for day 01/07/2012 across different optimization methods.}
\label{Profit_and_Time_Comparison}
\end{figure}

\begin{figure*}[htbp]
\centerline{\includegraphics[width=0.8\textwidth, trim=1.5cm 0.01cm 2.5cm 0.2cm, clip]{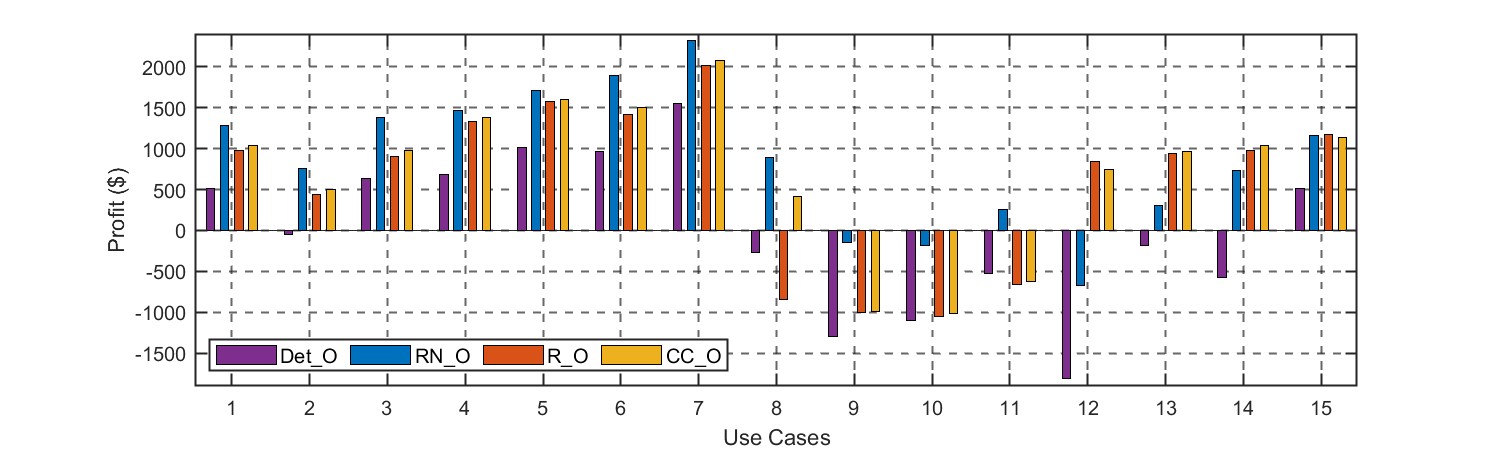}}
\caption{Test profits for day 01/07/2012 across different optimization methods under use cases of different contexts.}
\label{Context_New_Comparison}
\end{figure*}

The results in Fig.~\ref{profits_time_violations} indicate that the penalty-based LP relaxation provides a scalable alternative to the MILP reformulation when handling a large number of scenarios. While the MILP reformulation strictly enforces the BES complementarity condition and therefore results in zero violations, it becomes computationally expensive as the number of scenarios grows. In contrast, the penalty-based LP relaxation achieves nearly identical profit outcomes with significantly lower computational time, and the violations of the mutual exclusivity condition in BES charging and discharging decisions remain negligible. The McCormick relaxation also yields profit levels similar to both other methods; however, it exhibits substantially higher violations of the mutual exclusivity as the number of scenarios increases. Overall, the penalty-based LP relaxation offers the most practical trade-off between accuracy and scalability in the risk-neutral setting.

In case of \eqref{RO_1}, we define a box uncertainty set whose bounds are determined by the scenario set $S$ to solve the robust optimization problem. In contrast, for the chance-constrained problem \eqref{CC_1}, we initially selected $N_s = 194$ samples from the scenario set 
$S$ using the probabilistically robust approach \eqref{scenarios}, based on the desired values 
$\epsilon = 0.1$ and $\beta = 10^{-4}$, with $N_\xi = 2$. Then, we solve a standard robust problem using the bounds of the probabilistically selected set.

Fig.~\ref{Profit_and_Time_Comparison} shows the optimum profit and computational time for different optimization methods in the baseline case during training and testing. For brevity, these results are combined with those from the decision-making preference context, where the affine recourse policies are considered. We abbreviate the deterministic, risk-neutral, robust, and chance-constrained methods as `Det\_O', `RN\_O',  `R\_O', and `CC\_O', respectively, for clarity. From Fig.~\ref{Profit_and_Time_Comparison} (a), it is observed that the deterministic method produces the highest profit during training, since it ignores uncertainty. However, in testing, every realization results in mismatches, causing substantial ADC violations. Once penalties are applied, its profit drops sharply. Among the stochastic approaches, the chance-constrained method slightly outperforms the robust one in both training and testing, as it allows up to 10\% constraint violations in exchange for higher gains. The risk-neutral method stands out with the largest training profit among the stochastic strategies, as it optimizes over the expectation of all possible realizations. In this test case, the actual forecast errors were relatively small, resulting in zero ADC violations for all stochastic methods, which also enabled the risk-neutral method to secure a comparatively high test profit. It is also notable that the use of recourse policies does not significantly alter the profits of stochastic methods, while Fig.~\ref{Profit_and_Time_Comparison} (b) shows that they consistently reduce computation times in both training and testing. This benefit is minor for linear models, but could be far more valuable in nonlinear formulations (with network constraints), where computational complexity is much higher. 

In Fig.~\ref{Context_New_Comparison}, we compare the test performance of the optimization methods across several use cases with varying operational and market settings and uncertainty characteristics, as summarized in Table~\ref{context1}, where ``Low", ``Mod", and ``High" correspond to 50\%, 100\%, and 150\% of the original values, respectively. Case~1 serves as the baseline, with detailed results already discussed in Fig.~\ref{Profit_and_Time_Comparison}. Cases~1-2 confirm that joint participation in energy and FCAS markets substantially increases aggregator revenues across all methods, while Cases~3-4 highlight the sensitivity of profits to high FCAS MCPs. Note that Cases~1-11 maintain consistent treatment of uncertainty between the training and testing phases, ensuring the correct capturing of uncertainty while varying operational and market parameters. In Cases~1-7, the BES capacity is sufficient to absorb PV and load variations while concurrently providing energy and FCAS services. Consequently, all stochastic methods achieve zero ADC violations and demonstrate a consistent performance pattern with comparable positive profits to the baseline case. However, in Cases~8-11, where BES capacity is reduced compared to load and PV, particularly robust and chance-constrained methods exhibit significantly lower (mostly negative) profits due to their conservativeness: the BES must withhold flexibility for extreme realizations, limiting market participation. Negative profits (costs) mainly arise because insufficient BES capacity forces the aggregator to purchase power to meet demand. A notable exception is Case~8, where the chance-constrained method significantly outperforms the robust one due to reduced conservativeness and marginally sufficient flexibility offered by moderate BES and high PV capacities. Overall, across Cases~1–11 with varying operational and market settings and correctly captured uncertainty, the risk-neutral method consistently outperforms the others by optimizing over expected realizations, while the conservative nature of the robust and chance-constrained methods further reduces profits under limited BES capacity.

In Cases~12-15, uncertainty characteristics are varied between the training and testing phases to examine how inaccurate capturing of uncertainty affects the performance of stochastic methods. In Case~12, scaling the actual load by 25\% increases real-time power mismatches, resulting in significant profit losses for the risk-neutral method due to high ADC violations and associated penalties. The robust and chance-constrained methods outperform in this case, with the robust method showing the highest profit due to the lowest violations. In Case~13, where the penalty cost is lower, the impact of violations is reduced, allowing the chance-constrained method to slightly surpass the robust one. Case~14, which scales down actual PV realization, also increases real-time power mismatches and exhibits a similar performance pattern. Case~15 contracts the uncertainty set, leading all stochastic methods to produce comparable results, with the robust method marginally excelling due to its conservativeness.

\begin{figure}[htbp]
\centerline{\includegraphics[width=0.49\textwidth]{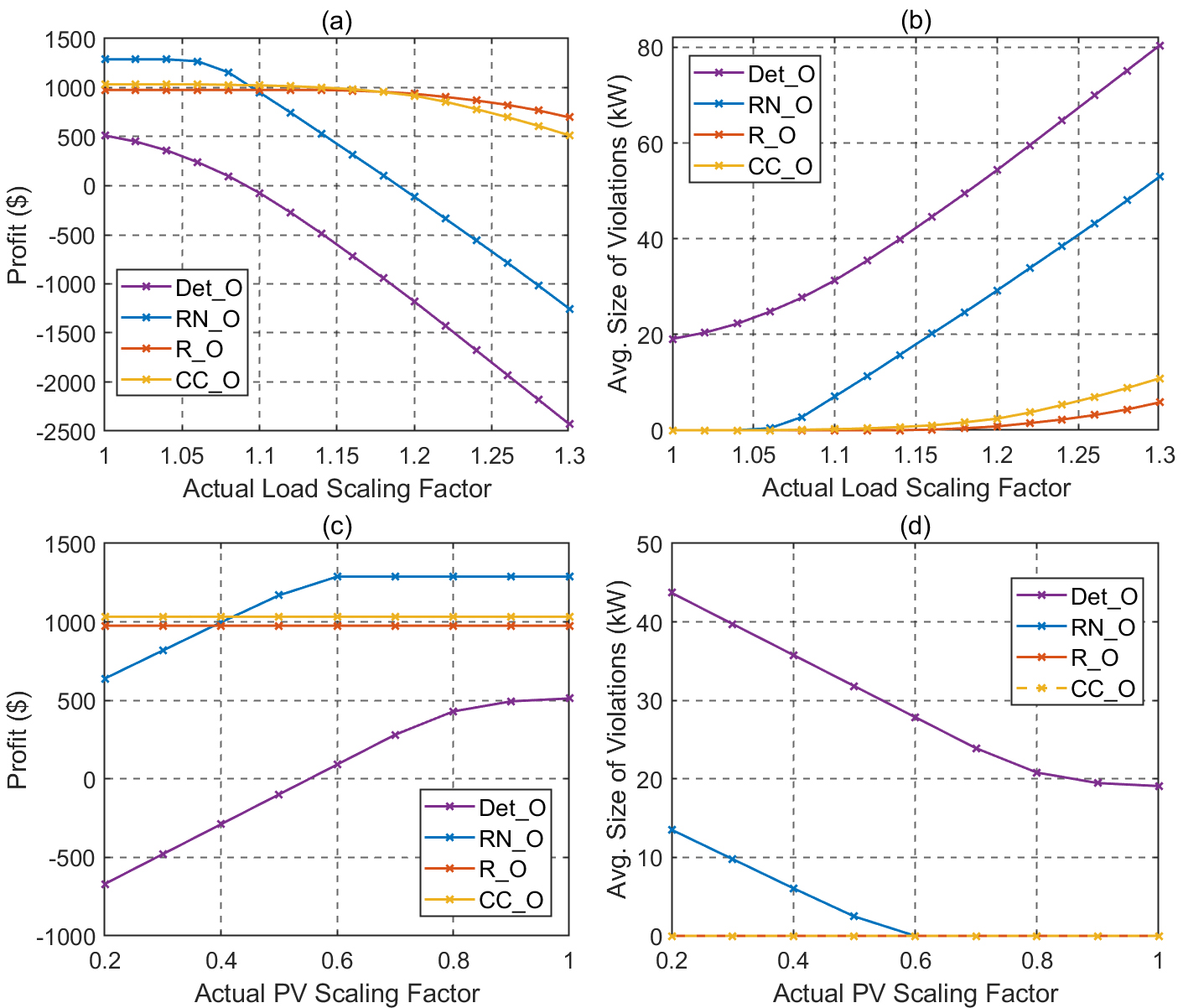}}
\caption{(a) Test profits and (b) average size of ADC violations, when the actual realization of the load is scaled up. (c) Test profits and (d) average size of ADC violations, when the actual realization of the PV is scaled down.}
\label{Actual_Load_and_PV_Scaling}
\end{figure}

Next, building on Cases~12 and 14, the actual load and PV realizations are scaled in finer steps to emulate varying real-time forecast errors, capturing how the performance varies among optimization methods and the point where the trade-off between risk-neutral and risk-averse methods emerges. The resulting profit and the average size of ADC violations for different optimization methods
on 01/07/2012, are shown in Fig.~\ref{Actual_Load_and_PV_Scaling}. Figure~\ref{Actual_Load_and_PV_Scaling}(a) shows that profits vary noticeably when the methods are tested under higher actual load realizations. As actual load levels rise, power mismatches generally increase, leading to more ADC violations across all optimization methods, as illustrated in Fig.~\ref{Actual_Load_and_PV_Scaling}(b). These mismatches trigger penalties, which in turn cause profits to decline. While the risk-neutral method maintains relatively higher profits under moderate conditions with limited violations, the robust approach becomes more advantageous as violations escalate with higher power mismatches. Similar patterns in profits and ADC violations appear under lower PV realizations in Fig.~\ref{Actual_Load_and_PV_Scaling}(c) and Fig.~\ref{Actual_Load_and_PV_Scaling}(d), but mainly for the deterministic and risk-neutral methods. Since PV output is comparatively smaller than the household load on this day, scaling it contributes little to total mismatches. As a result, the robust and chance-constrained methods remain unaffected, with no noticeable change in profit or ADC violations. Overall, the risk-neutral method’s profits fall below those of the other stochastic methods when the unseen power mismatch exceeds the expected mismatch by about 75\% on average.

\begin{figure}[htbp]
\centerline{\includegraphics[width=0.49\textwidth]{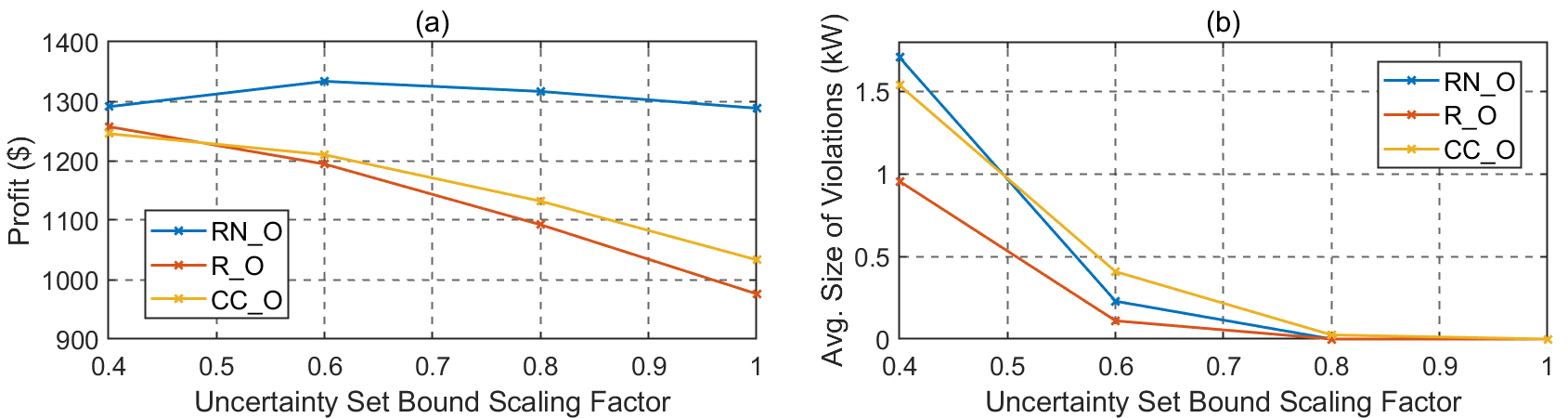}}
\caption{(a) Test profits and (b) average size of ADC violations, when the bound of the uncertainty set is scaled down.}
\label{Uncertainty_Set_Bound_Scaling_Profit_and_Violations}
\end{figure}

Next, extending Case~15, Fig.~\ref{Uncertainty_Set_Bound_Scaling_Profit_and_Violations} compares test profits and ADC violations under varying uncertainty set bounds. As the bounds contract, profits increase for all methods due to reduced conservativeness, with their values moving closer together at a 60\% contraction as shown in Fig.~\ref{Uncertainty_Set_Bound_Scaling_Profit_and_Violations}(a). However, as shown in Fig.~\ref{Uncertainty_Set_Bound_Scaling_Profit_and_Violations}(b), this contraction also raises ADC violations, notably for the risk-neutral method, since smaller scenario magnitudes in the training phase amplify power mismatches in the testing phase, thereby increasing penalties. Robust and chance-constrained methods also see more violations but remain more resilient than the risk-neutral approach.

\section{Conclusions} \label{C}

This paper presents three two-stage stochastic optimization methods: risk-neutral, robust, and chance-constrained, to guide CER aggregators in selecting context-aware bidding strategies for joint energy-FCAS markets under PV and load uncertainty. The formulations aim to ensure ADC, feasibility, and scalability, and are solved using scenario-based methodologies. Experiments have been conducted using real data under three different contexts: (i) operational and market settings; (ii) uncertainty characteristics; and (iii) decision-making preferences. 

In the first context, when uncertainty is correctly captured, the risk-neutral method is preferable as it consistently achieves the best performance across varying operational and market settings. The chance-constrained method typically follows, outperforming the robust approach by allowing limited (10\%) constraint violations. The choice of method becomes critical under limited BES capacity, as sufficient flexibility enables all stochastic methods to maintain zero ADC violations and stable profits, whereas smaller BES capacities reduce the effectiveness of the robust and chance-constrained methods due to their conservativeness.

In the second context, under varying uncertainty characteristics, the risk-neutral method is recommended when actual power mismatches are expected to remain small. When uncertainty is inaccurately captured or larger mismatches are likely, the robust and chance-constrained approaches become more suitable, with the robust method offering stronger protection due to its conservativeness. Practitioners should note that scaling the uncertainty set bounds exposes a clear profit-risk trade-off: contracting the bounds can improve profits by reducing conservativeness but increases exposure to penalties during testing, particularly for the risk-neutral method.

In the third context, decision-making preferences suggest that affine recourse policies are recommended for all stochastic methods when faster solution times are required, as they significantly reduce computation effort with only marginal impacts on profit.

Overall, method selection under uncertainty should be context-aware: risk-neutral methods are recommended when uncertainty is accurately captured under typical operational and market conditions; robust methods are more suitable under high uncertainty, particularly when large power mismatches arise from inaccurately modeled uncertainty; and chance-constrained methods are preferred when seeking a balance between profitability and controlled risk of constraint violations.

\vspace{-0.2cm}
\bibliographystyle{ieeetr}
\bibliography{references}

\end{document}